**Y. Yao, Y. Tsusaka, and Y. Ishikawa**

**Kinematical and dynamical contrast of dislocations in thick GaN substrates observed by synchrotron-radiation x-ray topography under six-beam diffraction conditions,**

# Kinematical and dynamical contrast of dislocations in thick GaN substrates observed by synchrotron-radiation X-ray topography under six-beam diffraction conditions

Yongzhao Yao,[1,2,a)] Yoshiyuki Tsusaka,[3] and Yukari Ishikawa[2]

[1]Innovation Center for Semiconductor and Digital Future, Mie University, 1577 Kurimamachiya-cho, Tsu, Mie 514-8507, Japan

[2]Japan Fine Ceramics Center, 2-4-1 Mutsuno, Atsuta, Nagoya 456-8587, Japan

[3]Graduate School of Science, University of Hyogo, 3-2-1, Koto, Kamigori-cho, Ako-gun, Hyogo 678-1297, Japan

**Abstract:** Dislocations in a thick ammonothermal GaN substrate were investigated by synchrotron-radiation X-ray topography (SR-XRT) performed under six-beam diffraction conditions. The use of high-brilliance synchrotron radiation enabled the observation of the super-Borrmann effect, which significantly enhanced the anomalous transmission of X-rays through the 350-μm-thick crystal. The evolution of dislocation contrast with varying deviation angle $\Delta\omega$ revealed a transition from kinematical to dynamical diffraction, in agreement with theoretical predictions based on the dynamical theory. By selectively generating five equivalent two-beam diffraction conditions near the six-beam configuration, the Burgers vectors of individual threading edge dislocations (TEDs) were identified according to the $\boldsymbol{g}\cdot\boldsymbol{b}$ invisibility criterion. The measured line widths of the dislocation images were consistent with calculated values derived from the extinction distance and $|\boldsymbol{g}\cdot\boldsymbol{b}|$ dependence, confirming that most dislocations had a Burgers vector containing a $\boldsymbol{a}$-type component of $\frac{1}{3}\langle 11\bar{2}0\rangle$ or $\frac{2}{3}\langle 11\bar{2}0\rangle$. These results demonstrate that SR-XRT under multi-beam diffraction provides a powerful nondestructive approach for quantitative dislocation analysis in thick GaN crystals, offering valuable insight into defect structures relevant to high-performance GaN-based electronic devices.

[a)]Author to whom correspondence should be addressed. E-mail: yao@icsdf.mie-u.ac.jp. ORCID:0000-0002-7746-4204

**Authors' ORCID and E-mail address**

| | | |
|---|---|---|
| **Yongzhao Yao** | **0000-0002-7746-4204** | **yao@icsdf.mie-u.ac.jp** |
| **Yoshiyuki Tsusaka** | **0000-0001-7123-1269** | **tsusaka@sci.u-hyogo.ac.jp** |
| **Yukari Ishikawa** | **0000-0001-8908-7592** | **yukari@jfcc.or.jp** |

# I. INTRODUCTION

Gallium nitride (GaN) has attracted extensive attention as a key material for next-generation high-frequency and high-power electronic devices due to its excellent physical properties [1]. Despite these advantages, the growth of high-quality and large-diameter bulk GaN crystals remains a significant challenge. Currently, commercial GaN substrates are primarily produced by hydride vapor phase epitaxy (HVPE) [2, 3] and ammonothermal methods [4, 5, 6], while Na-flux growth is under active development in academic laboratories [7]. For GaN-based power devices, crystalline imperfections critically limit device performance. In particular, threading dislocations act as leakage paths and reduce the breakdown voltage, resulting in device degradation and premature failure [8, 9, 10]. Therefore, the detection and characterization of dislocations in GaN are of great importance, especially by nondestructive methods that can be applied to substrates before device fabrication.

X-ray topography (XRT) has proven to be a powerful tool for the nondestructive evaluation of crystalline defects, and XRT observations with both reflection (Bragg case) [11, 12, 13, 14, 15, 16, 17, 18] and transmission (Laue case) [5, 19] configurations have been applied to GaN crystals. However, direct observation of dislocations in the deep region (>100 μm) of thick GaN substrates remains difficult due to the strong X-ray absorption arising from the presence of heavy Ga atoms. Obtaining transmission topography images of GaN based on kinematical contrast would require a very thin crystal (< 100 μm) to satisfy the $\mu t \sim 1$ condition, where $\mu$ is the linear absorption coefficient and $t$ is the sample thickness [19]. To overcome this limitation, synchrotron-radiation (SR) XRT observation performed under the Borrmann effect [20, 21, 22], a dynamical X-ray diffraction phenomenon also known as anomalous transmission, provides a promising approach [5], as it enables enhanced transmission of X-rays through perfect crystal regions while strongly revealing distortions around dislocations [23, 24, 25]. For a highly perfect and thick [20] crystal oriented for Laue geometry diffraction, the two plane waves corresponding to the primary and diffracted X-rays, respectively, are coherently coupled in the crystal, and their interference produces a set of standing waves. When the zero-amplitude points (nodes) of the standing waves coincide with the atomic planes (i.e., the Bragg reflecting planes), photoelectric absorption is significantly reduced, leading to a marked increase in transmitted X-ray intensity. This is known as anomalous transmission. Moreover, under multiple-beam ($n$-beam) diffraction conditions, the effective absorption coefficient can be further reduced due to the interference of the primary wave with multiple diffracting waves, thus enabling the visualization of dislocations in much thicker GaN crystals than was previously possible. This further-enhanced Borrmann effect is called the super-Borrmann effect [26], which has been applied to Ge [27] and Si [28] using three-beam diffraction to observe dislocations in thick bulk crystals. The dynamical images of lattice distortion observed under $n$-beam diffraction have been theoretically and experimentally studied by several groups based on a generalized Takagi–Taupin theory [29, 30, 31, 32, 33, 34, 35].

In this work, we performed SR-XRT observations under six-beam diffraction to reveal dislocations over a wide area of a thick ammonothermal GaN substrate. We demonstrate that Burgers vectors can be analyzed using the $\boldsymbol{g}\cdot\boldsymbol{b}$ invisibility criterion (where $\boldsymbol{g}$ is the diffracting vector in reciprocal space and $\boldsymbol{b}$ is the Burgers vector of a dislocation) [36], similar to analyses based on kinematical contrast [19].

## II. EXPERIMENTAL DETAILS

SR-XRT measurements were performed at beamline BL24XU of SPring-8, a third-generation synchrotron facility that provides an X-ray source with high brilliance and low emittance. Such a source can be regarded as a plane wave, thereby enabling the observation of dynamical X-ray diffraction in GaN under six-beam conditions. Beamline BL24XU delivers a 1.2 mm × 1.2 mm undulator beam with an energy resolution of ΔE/E better than $10^{-4}$ and a beam divergence of approximately 5 arcsec [37]. The X-ray wavelength was set to 1.24 Å (E = 10 keV), which lies slightly on the low-energy side of the Ga K-edge at 1.20 Å [24, 38]. The experimental setup is schematically illustrated in Fig. 1(a) and was similar to those employed for studies of other semiconductor materials at this beamline [19, 24, 25, 27, 39, 40, 41]. A commercial 1-inch-diameter ammonothermal GaN single-crystal substrate [5] was examined. The substrate has a {0001}-oriented surface. Both sides of the substrate were subjected to chemical–mechanical polishing to remove surface damage that could otherwise hinder XRT observation. The X-ray entrance and exit surfaces of the substrate correspond to the N-polar and Ga-polar faces, respectively. The substrate thickness was 350 μm, corresponding to a $\mu t$ value of 5.5. Since the transmitted X-ray intensity is proportional to $e^{-\mu t}$, the normal transmission intensity without the Borrmann effect is only about 1% of that under the typical $\mu t = 1$ condition for transmission XRT. This is consistent with our observation that the transmitted beam produced only a faint Laue spot on the fluorescent screen when the sample configuration deviated from an $n$-beam diffraction condition ($n > 1$).

The sample configuration was adjusted to the six-beam condition by controlling the φ, ω, and ψ angles as follows (φ: rotation about the surface normal; ω and ψ are defined in Fig. 1(a)). First, using the orientation flat of the substrate as a reference, the sample was mounted on the holder with φ carefully set so that the $[01\bar{1}0]$ direction was vertical (upwards). In other words, the $m$-planes, i.e., the $(01\bar{1}0)$ planes, were oriented horizontally when ω = 0°. At ω = 0°, the surface normal of the sample also lay in the horizontal plane. Next, ω was tilted by about 27° to satisfy the Bragg condition for $\boldsymbol{g}$ =

$02\bar{2}0$ (denoted $\boldsymbol{g}_3$ in **Fig. 1(c)**). At this orientation, two strong Laue spots appeared on the fluorescent screen, corresponding to the transmitted beam (o-wave) and the $02\bar{2}0$ diffracted beam. In this diffraction geometry, the scattering plane is oriented vertically. This condition allows the Borrmann effect to occur under two-beam diffraction. At this stage, if the $\boldsymbol{g}_3$ spot on the fluorescent screen was not exactly aligned above the o spot, φ was further adjusted until perfect alignment was achieved. After φ and ω were set, the third axis, ψ, was adjusted. Rotation about ψ is defined with respect to the $\boldsymbol{g}_3$ direction; therefore, the $\boldsymbol{g}_3$ spot remains bright during this rotation, since its reciprocal lattice point is invariant on the Ewald sphere. By carefully tuning ψ such that the plane containing the reciprocal lattice points $\boldsymbol{g}_1$–$\boldsymbol{g}_5$ intersects the Ewald sphere, the six-beam diffraction condition was established. Under this condition, six Laue spots with comparable brightness appeared on the fluorescent screen, as shown in **Fig. 1(b)**. The optical paths are illustrated in **Fig. 1(c)**. The directions of the wavevectors $\boldsymbol{K}_0$~$\boldsymbol{K}_5$, corresponding to the o-wave and the five diffracted waves $\boldsymbol{g}_1$–$\boldsymbol{g}_5$, are indicated in **Fig. 1(a)**. The six-beam diffraction condition can always be achieved by adjusting the three angles, owing to the fact that GaN has a hexagonal crystal structure belonging to the space group $P6_3mc$. Consequently, the reciprocal lattice points $\boldsymbol{g}_1$–$\boldsymbol{g}_5$ together with the o-spot form a regular hexagon (**Fig. 1(d)**), which can intersect with the Ewald sphere regardless of the wavelength, provided that the angles are properly adjusted [**41**]. This situation is similar to six-beam diffraction in Si [**32**], although Si has a diamond crystal structure.

After the sample was oriented for the six-beam diffraction condition, the fluorescent screen was moved upward to allow the forward-transmitted beam to enter the imaging system (**Fig. 1(a)**), which consists of a scintillator, relay lenses, and a CMOS detector. The relay lenses provide a 10× magnification. Given that the pixel size of the CMOS detector is 6.5 μm, the imaging system achieves an effective spatial resolution of 0.65 μm/pixel [**42**].

## III. RESULTS AND DISCUSSION

Fig. 2 shows the intensity of the forward-transmitted beam (o-wave) as a function of $\Delta\omega$, representing the deviation from the six-beam diffraction condition at $\omega \approx 27°$. The intensity was obtained by integrating all pixels in the XRT images recorded at each $\omega$ value. Representative XRT images are shown at the maximum intensity (exact six-beam condition) and at the half-maximum intensities on both sides of the peak. At the six-beam diffraction condition, the reciprocal-lattice points of o and $\boldsymbol{g}_1$–$\boldsymbol{g}_5$ intersect the Ewald sphere. When $\Delta\omega < 0$, $\boldsymbol{g}_1$–$\boldsymbol{g}_5$ lie outside the Ewald sphere, whereas for $\Delta\omega > 0$ they lie inside it, corresponding to negative and positive excitation errors, respectively. The asymmetry observed in the $\omega$ rocking curve of the transmitted intensity near the six-beam diffraction condition is a characteristic feature arising from the dynamical diffraction process. [21, 22, 43].

The three XRT images in Fig. 2 reveal the presence of dislocations in the sample. Most of these dislocations are attributed to threading dislocations running parallel to the *c*-axis of the crystal. Although the shape and fine details of the dislocation contrast differ among the images, a one-to-one correspondence can be established between the positions of individual dislocations. In particular, the upper end of each dislocation image is well aligned across the images. From Fig. 1(a), it is evident that the upper end of the dislocation in the XRT images corresponds to the dislocation termination point at the X-ray exit surface. This observation is understandable because the lattice distortion near the exit surface is projected in the XRT image without significant in-plane shift, as it is located near the base of the Borrmann triangle. In contrast, lattice distortion near the entrance surface generates a new wavefield that interferes with the primary wavefield, causing the corresponding dislocation images to spread over the entire Borrmann triangle.

More details of the dislocation contrast as a function of Δω are presented in Fig. 3. The images cover the range of Δω from −0.007° to 0.003° with an interval of 0.001°, where the exact six-beam diffraction condition corresponds to Fig. 3(h). Image brightness and contrast were adjusted for better viewing. When Δω < 0, the dislocations appear as straight lines. This type of dislocation contrast mainly arises from kinematical diffraction and is referred to as the "direct image" in Authier's textbook [22]. When the ω value deviates significantly from the six-beam diffraction condition, all five reciprocal space points $\boldsymbol{g}_1$–$\boldsymbol{g}_5$ lie at a considerable distance from the Ewald sphere. In this case, only strongly distorted regions, where the $\boldsymbol{g}$ point is largely displaced, intersect the Ewald sphere—such as the areas near the dislocation core—whereas the perfect regions do not satisfy the Bragg condition. Distorted regions in the immediate vicinity of the dislocations give rise to kinematical diffraction, and as a consequence, the dislocation contrast manifests as straight, thin lines. The relatively thick line in the lower-left part of the images (marked by a red arrow) is attributed to a dislocation with a larger Burgers vector.

It should be noted that, overall, the line width increases as Δω approaches zero, indicating an increasing contribution of dynamical diffraction relative to kinematical diffraction. When the deviation of $\boldsymbol{g}_1$–$\boldsymbol{g}_5$ becomes sufficiently small so that they nearly intersect the Ewald sphere, the dynamical mechanism becomes dominant. Under this condition, the dislocation contrast is transformed from straight lines into a triangular form, as shown in Fig. 3(g)–(i). The apex of the triangle represents the termination point of the dislocation at the exit surface, while the base reflects the lateral spread of the dislocation over the Borrmann triangle [21, 22, 43]. Another characteristic feature of dynamical diffraction can then be observed: the pendellösung fringes around the dislocation line. The pendellösung fringes, which can be interpreted as a beating pattern of the X-rays, arise from the interference between the primary wave and the new wavefield generated at the dislocation, and their periodicity corresponds to the extinction distance of the X-rays.

With further increases in Δω on the Δω > 0 side, as shown in **Fig. 3(j)** and **3(k)**, the dynamical effect is weakened. Under this condition, the distorted regions surrounding the dislocation cores fail to satisfy the Bragg condition. Consequently, the dislocation lines evident in **Fig. 3(a)–(f)** disappear, and only the termination points of the dislocations manifest as spot-like contrasts in the images. Six-beam diffraction XRT images acquired over a wider area and with a finer ω interval of 0.0004° are presented in **Supplementary Figure 1**.

The dislocation contrast observed under the six-beam diffraction condition in our work agrees well with the dynamical theory presented by A. Authier [22] and with simulation results based on numerical integration of the Takagi–Taupin equations, as reported by Suvorov *et al.* [43, 44, 45, 46, 47, 48] and by Epelboin and co-workers [49, 50, 51, 52]. These studies show that a dislocation image consists of three parts: (1) the "direct" or kinematical image, which is formed in the strongly distorted region of the dislocation elastic field and provides accurate information about the actual position of the dislocation; (2) the dynamical image, which arises from wavefield redistribution in the Borrmann triangle and appears as a light shadow in the topograph; and (3) the intermediate image, which results from the interference between the wavefield propagating in the Borrmann triangle and the new wavefields generated in the strongly distorted region near the dislocation core [43].

**Fig. 4** shows an XRT image obtained under the six-beam diffraction condition from a wide area of 3.5 mm × 3.4 mm near the substrate center. The image was constructed by tiling 7 × 7 sub-images, each recorded by translating the sample in the X–Y direction parallel to the sample surface. Dislocations, most of them threading dislocations, are clearly visible. As explained above, the dislocation contrast consists of direct, dynamical, and intermediate components, and the line direction of the dislocations can be determined from the direct contrast. It was found that not all dislocations share the same line direction; instead, they deviate from the exact c-axis. The degree of deviation and tilt angle depends on their position relative to the substrate center. This behavior is considered to result from the thermal distribution in the growth chamber and/or the shape of the bulk crystal growth front, which affects the propagation direction of the threading dislocations [**4**, **5**, **6**, **13**, **53**, **54**].

An important task in the dislocation observation of GaN is to identify the Burgers vectors of the dislocations. In the present experimental setup, this can be readily achieved by generating five two-beam diffraction conditions (i.e., within the two-beam approximation) in the vicinity of the six-beam diffraction condition. First, the sample position was adjusted to satisfy the six-beam diffraction condition, as explained above. Under this condition, the six reciprocal spots o and $\boldsymbol{g}_1$–$\boldsymbol{g}_5$ lie exactly on the surface of the Ewald sphere (**Fig. 5(a)**). Starting from this alignment, the azimuthal angle ψ was rotated so that $\boldsymbol{g}_3$ remained on the Ewald sphere (i.e., retaining its brightness), while $\boldsymbol{g}_1$ and $\boldsymbol{g}_2$ were rotated out of the Ewald sphere and $\boldsymbol{g}_4$ and $\boldsymbol{g}_5$ were rotated into the Ewald sphere (**Fig. 5(b)–(e)**). After ψ was rotated by 0.01°, at which point $\boldsymbol{g}_1$, $\boldsymbol{g}_2$, $\boldsymbol{g}_4$, and $\boldsymbol{g}_5$ appeared faint on the fluorescent screen, a rocking scan of the ω angle was performed. The purpose of this ω rocking was to bring each of the $\boldsymbol{g}_1$–$\boldsymbol{g}_5$ spots individually back onto the surface of the Ewald sphere, thereby creating five independent two-beam diffraction conditions corresponding to $\boldsymbol{g}_i$ (i = 1–5). Notably, the total rocking range was only 0.1°, ensuring that there was almost no image distortion among the XRT images obtained under the five two-beam diffraction conditions. This enabled direct comparison of the XRT contrast of the same dislocations across the five images. A movie demonstrating how the five $\boldsymbol{g}$ spots were sequentially brought onto the Ewald sphere is provided as **Supplementary Movie 1**.

During the ω rocking, XRT images were continuously recorded using the o-wave. When each $\boldsymbol{g}$ spot intersected the Ewald sphere, the XRT image corresponding to that specific $\boldsymbol{g}$-vector was obtained (see **Supplementary Movie 2**), enabling $\boldsymbol{g}\cdot\boldsymbol{b}$ invisibility analysis for identifying the Burgers vectors of dislocations. **Fig. 6** shows the integral intensity of the o-wave as a function of Δω at ψ = 0.01°. Five XRT images corresponding to $\boldsymbol{g}_1$–$\boldsymbol{g}_5$ are also presented. Among the five peaks, the intensity follows the order $\boldsymbol{g}_2 \approx \boldsymbol{g}_4 > \boldsymbol{g}_1 \approx \boldsymbol{g}_5 > \boldsymbol{g}_3$. This trend arises from differences in the structure factors of the $11\bar{2}0$, $01\bar{1}0$, and $02\bar{2}0$ families, as well as from the varying strength of the Borrmann effect for different sets of atomic planes. From the XRT images, we found that the ratio of kinematical to dynamical contrast is higher for $\boldsymbol{g}_1$, $\boldsymbol{g}_3$, and $\boldsymbol{g}_5$ than for $\boldsymbol{g}_2$ and $\boldsymbol{g}_4$. This is evident because the images for $\boldsymbol{g}_1$, $\boldsymbol{g}_3$, and $\boldsymbol{g}_5$ clearly show dislocation lines as direct contrast, whereas those for $\boldsymbol{g}_2$ and $\boldsymbol{g}_4$ display only the termination points of dislocations. Notably, direct contrast also appears at $\boldsymbol{g}_2$ and $\boldsymbol{g}_4$ when a small deviation from the exact Bragg condition is introduced. Under such conditions, dislocation lines become clearly visible (see **Supplementary Movie 2**).

It is noteworthy that, due to differences in the direction of the $\boldsymbol{g}$-vectors, the Borrmann shadow of a threading dislocation appears on opposite sides of the direct dislocation image. For example, in the $\boldsymbol{g}_1$ image, the threading dislocations cast a triangular dynamical contrast on the right side of the dislocation lines, whereas in the $\boldsymbol{g}_5$ image, the triangular contrast appears on the left. Interestingly, the Borrmann shadow is not evident in the $\boldsymbol{g}_3$ image, because the dislocation line lies within the scattering plane. This observation is consistent with the series of simulation studies by Suvorov *et al*. [**43**, **44**, **45**, **46**], who, using numerical calculations based on the Takagi–Taupin equations, demonstrated how dynamical dislocation contrast varies with the dislocation line direction, the Burgers vector, the applied $\boldsymbol{g}$-vector, and the depth of the dislocation relative to the entrance and exit surfaces.

Using the five sets of images acquired under two-beam diffraction conditions with $\boldsymbol{g}_1$–$\boldsymbol{g}_5$, the Burgers vectors were evaluated based on the $\boldsymbol{g}\cdot\boldsymbol{b}$ invisibility criterion. **Fig. 7(a)–(c)** present an example in which the same region was observed with three $1\bar{1}00$-type $\boldsymbol{g}$-vectors. These images were extracted from a sequence of continuously recorded frames, as shown in **Supplementary Movie 2**. Since the in-plane Burgers vectors in GaN typically have a $\frac{1}{3}\langle 11\bar{2}0\rangle$ component corresponding to the lattice constant $\boldsymbol{a}$ = 3.189 Å, a threading edge dislocation (TED) should be out of contrast under one of the $\boldsymbol{g}_1$, $\boldsymbol{g}_3$, or $\boldsymbol{g}_5$ conditions. In **Fig. 7**, the two dislocations labeled 1 and 2 (indicated by red arrows) possess Burgers vectors perpendicular to $\boldsymbol{g}_1$, as they are clearly visible in **Figs. 7(b) and 7(c)** but nearly invisible in **Fig. 7(a)**. Here, "nearly invisible" refers to a very weak residual contrast that arises when the condition $\boldsymbol{g}\cdot\boldsymbol{b}$ = 0 is satisfied but $\boldsymbol{g}\cdot(\boldsymbol{b}\times\boldsymbol{l})$ = 0 is not, where $\boldsymbol{l}$ denotes the direction vector of the dislocation line [36]. Similarly, the dislocation labeled 3 (indicated by a green arrow) has a Burgers vector perpendicular to $\boldsymbol{g}_3$, being visible in **Figs. 7(a) and 7(c)** but nearly invisible in **Fig. 7(b)**. Similar behavior was observed for dislocations 4 and 5, whose Burgers vectors are perpendicular to $\boldsymbol{g}_5$. It is noted that no $\boldsymbol{m}$-type threading edge dislocation (TED) with a Burgers vector of pure $\langle 1\bar{1}00\rangle$ type was found in this sample, although such TEDs have been reported in 4H-SiC [55]. An $\boldsymbol{m}$-type TED would be invisible under $\boldsymbol{a}$-type $\boldsymbol{g}$-vectors such as $\boldsymbol{g}_2$ and $\boldsymbol{g}_4$. However, no dislocation invisibility was observed in these two images when compared with those taken under $\boldsymbol{g}_1$, $\boldsymbol{g}_3$, and $\boldsymbol{g}_5$.

It is important to point out that, in theory, a pure threading screw dislocation (TSD) with a Burgers vector of [0001] or [000$\bar{1}$] cannot be detected under $\boldsymbol{g}_1$–$\boldsymbol{g}_5$, because both $\boldsymbol{g}\cdot\boldsymbol{b}$ = 0 and $\boldsymbol{g}\cdot(\boldsymbol{b}\times\boldsymbol{l})$ = 0 are always satisfied. Nevertheless, a very weak contrast can still be expected due to the Eshelby twist [56, 57, 58, 59, 60, 61] appearing at the surface termination points of the TSDs. Consequently, a TSD would exhibit a faint Eshelby-type contrast in all five images corresponding to $\boldsymbol{g}_1$–$\boldsymbol{g}_5$. No dislocation exhibiting such features was identified in the present observations. We believe this is because the GaN crystal grown by the current method contains very few TSDs. This interpretation is consistent with previous results obtained from reflection XRT by Sintonen *et al.* [11, 12] and Yao *et al.* [14]. Since the $\boldsymbol{g}$-vectors $\boldsymbol{g}_1$–$\boldsymbol{g}_5$ are insensitive to the ⟨0001⟩ component, it is appropriate to note that the five dislocations discussed above in Fig. 7 may be threading mixed dislocations (TMDs) possessing both $\boldsymbol{c}$ and $\boldsymbol{a}$ components.

Next, the phenomenon of the super-Borrmann effect and its influence on dislocation contrast are discussed. Suvorov *et al.* [48] demonstrated that, under two-beam diffraction conditions, interbranch scattering plays a dominant role in the formation of dislocation contrast. This interbranch scattering is associated with the transfer of energy from the abnormally transmitted wave to the abnormally absorbed wave (i.e., the highly absorbed wave) in the vicinity of a dislocation, caused by the local lattice distortion. By reasonably simplifying the distortion field around a dislocation, the authors were able to describe the X-ray wave field within the crystal using a set of ordinary differential equations analogous to the Howie–Whelan equations widely employed for interpreting image contrast in transmission electron microscopy [62], instead of the more complex Takagi–Taupin partial differential equations [22]. In this formulation, the overall effective absorption coefficient ($\mu_{\mathrm{eff}}$) of the crystal can be expressed as an integral of the local effective absorption coefficient along the trajectory of the abnormally transmitted Bloch wave throughout the entire crystal thickness ($t$) [48]:

$$\mu_{eff} = \mu t^{-1} \int_0^t \left(1 - \frac{1}{\sqrt{1+\beta^2}}\right) dz \qquad \text{... (Eq. 1)}$$

where $z$ denotes the coordinate along the crystal thickness direction, and $\beta$ is defined as a function of the extinction distance $\Lambda$ and the derivative, with respect to $z$, of the inner product between the $\boldsymbol{g}$-vector and the displacement field $\boldsymbol{u}$ associated with the dislocation:

$$\beta = \frac{1}{\Lambda} \cdot \frac{d(\boldsymbol{g} \cdot \boldsymbol{u})}{dz} \qquad \text{... (Eq. 2)}$$

$$\Lambda = \frac{\pi V \sqrt{\gamma_0 \gamma_g}}{r_e |C| \lambda \sqrt{F_g F_{\bar{g}}}} \qquad \text{... (Eq. 3)}$$

Here, $V$ is the volume of the unit cell, $\boldsymbol{g}$ is the diffraction vector, $\gamma_0$ and $\gamma_g$ are the cosines of the angles between the crystal surface normal and the incident and diffracted beam directions, respectively, $r_e$ is the classical electron radius, $C$ is the polarization factor, and $F_g$ is the structure factor [63, 64].

From **Eq. 1**, several important conclusions can be drawn regarding the dislocation contrast. First, in a highly perfect crystal far from any dislocation, $\mu_{eff}/\mu$ is much smaller than unity and approaches $\beta^2/2$; that is, the effective absorption coefficient is greatly suppressed relative to the normal photoelectric absorption coefficient. This occurs because the displacement field $\boldsymbol{u}$ is small, and consequently, $\beta$ is also small. Secondly, even in the vicinity of a dislocation, if $\boldsymbol{g}\cdot\boldsymbol{u} = 0$, that is, when a $\boldsymbol{g}$-vector perpendicular to the displacement field is used, $\beta$ remains small, and the transmitted intensity approaches that in a perfect crystal. In other words, the presence of the dislocation does not affect the transmitted intensity, and thus the dislocation becomes invisible. This observation justifies the use of the $\boldsymbol{g}\cdot\boldsymbol{b}$ invisibility criterion to identify the Burgers vector of a dislocation under the two-beam Borrmann condition, as explained above in **Fig. 7**. Thirdly, near a dislocation where $\boldsymbol{g}\cdot\boldsymbol{u} \neq 0$, $\beta$ becomes large, and therefore $\mu_{eff}/\mu$ approaches unity. In this case, no anomalous transmission occurs in the vicinity of the dislocation, and this region of the crystal exhibits normal photoelectric absorption to X-rays. As a result, the dislocation contrast appears against the background of anomalous transmission.

Furthermore, according to **Eq. 2**, it is evident that for a given $\boldsymbol{g}$-vector, whether the parameter β is small or large is determined by the strain gradient d$\boldsymbol{u}$/d$z$ relative to the extinction distance $\Lambda$. In the close vicinity of the dislocation core, the gradient of the lattice distortion becomes so large that the wavefield does not have sufficient time to adjust to the rapid lattice changes. As a result, the dynamical diffraction contrast breaks down, and only kinematical contrast is observed. This is consistent with the interpretation provided by Suvorov *et al*. [**48**], who explained the phenomenon in terms of interbranch scattering, suggesting that Bloch waves belonging to one sheet of the dispersion surface can generate waves corresponding to another sheet [**65**, **66**]. The kinematically scattered area ($S$) around the dislocation line can be estimated as [**63**]:

$$S = \frac{|\boldsymbol{g} \cdot \boldsymbol{b}|\Lambda^2 \cos(\theta)}{2\sqrt{2}\pi} \qquad \ldots(\text{Eq. } 4)$$

where $\theta$ is the Bragg angle. By substituting $F_g$ = 38 (with $f_{Ga}$ = 31 and $f_N$ = 7) [**67**], $r_e$ = 2.818×10$^{-5}$ Å, and $V$ = 45.67 Å$^3$ into **Eq. 3**, we obtain an extinction distance of $\Lambda$ = 10.5 μm for $\boldsymbol{g}$ = $1\bar{1}00$. Substituting this value into **Eq. 4** yields $S$ = 12.2 μm$^2$. Therefore, the line width of a TED with b = $\frac{a}{3}\langle 11\bar{2}0\rangle$ under kinematical diffraction at $\boldsymbol{g}$ = $1\bar{1}00$ is estimated to be approximately 3.93 μm.

**Fig. 8** shows the line width of two dislocations observed in **Fig. 3(a)** under kinematical diffraction with a large deviation (Δω) from the exact six-beam diffraction condition. This image can be regarded as a superimposition of five kinematical images, each corresponding to one of the reflections $\boldsymbol{g}_1$–$\boldsymbol{g}_5$. The line widths were determined by Gaussian fitting of the pixel intensity profiles shown in **Fig. 8(a) and (b)**, yielding $W_1$ =

8.11 μm and $W_2$ = 4.15 μm. The value of $W_2$ = 4.15 μm is in good agreement with the estimated width of 3.93 μm for a TED, showing a small deviation of approximately 6%. In this region, a large fraction of dislocations exhibit line widths similar to $W_2$, indicating the dominant presence of TEDs. On the other hand, from **Eq. 4**, we know that the line width $W$ is proportional to $\sqrt{|\boldsymbol{g} \cdot \boldsymbol{b}|}$. The measured value of $W_1$ = 8.11 μm agrees well the calculated width of 7.86 μm for $|\boldsymbol{g} \cdot \boldsymbol{b}| = 4$. The values of $|\boldsymbol{g} \cdot \boldsymbol{b}|$ for TEDs corresponding to $\boldsymbol{g}_1$–$\boldsymbol{g}_5$ are listed in **Table I**. The definitions of the Burgers vector and $\boldsymbol{g}$-vector directions are illustrated in **Fig. 9**. The table is calculated under the assumption that the TEDs possess an elementary Burgers vector of the $\boldsymbol{a}$-type $\frac{1}{3}\langle 11\bar{2}0\rangle$. Therefore, for a TED with $\boldsymbol{b} = 2\boldsymbol{a}$, namely, $\frac{2}{3}\langle 11\bar{2}0\rangle$, the corresponding $\boldsymbol{g}$ and $\boldsymbol{b}$ combinations in **Table I** with a value of 2 yields $|\boldsymbol{g} \cdot \boldsymbol{b}| = 4$. Threading dislocations in ammonothermal GaN with an in-plane Burgers vector of $2\boldsymbol{a}$ have been reported by Kanechika *et al.* [**16**].

In the case of anomalous transmission XRT performed under a pure two-beam diffraction condition that is far from any $n$-beam conditions, only the transmitted wave and one diffracted wave are excited. Consequently, all the transmitted radiation energy is confined within a triangular region known as the Borrmann triangle or Borrmann fan. In contrast, under six-beam diffraction conditions, the radiation energy is concentrated within a pyramid whose base is a hexagon formed by the reciprocal lattice vectors $\boldsymbol{g}_1$ through $\boldsymbol{g}_5$ and the origin O (**Fig. 1(d)**), and whose apex lies at the source point [**33**]. Simultaneous $n$-beam diffraction was theoretically studied by Joko *et al.* [**29**], who demonstrated that when the relevant reciprocal lattice points form symmetric configurations, such as a square ($n$ = 4) or a regular hexagon ($n$ = 6), as in the case of wurtzite GaN examined in this study, the minimum absorption coefficient is reduced compared to the two-beam case. The authors pointed out that a strong reduction of the absorption coefficient takes place in the case of exact multiple diffraction, where all the Bragg conditions are satisfied strictly. Therefore, six-beam diffraction requires extremely high crystal perfection to be observed.

## IV. CONCLUSIONS

Synchrotron-radiation XRT under six-beam diffraction conditions was successfully applied to visualize and analyze dislocations in a thick ammonothermal GaN substrate. The enhanced anomalous transmission due to the super-Borrmann effect enabled the observation of dislocations deep inside the 350-μm-thick crystal, which would otherwise be opaque under conventional transmission topography. The evolution of dislocation contrast with varying deviation angles (Δω) was systematically investigated and found to be in good agreement with dynamical diffraction theory, including the transition from kinematical to dynamical regimes and the appearance of pendellösung fringes. By generating five equivalent two-beam diffraction conditions ($\boldsymbol{g}_1$–$\boldsymbol{g}_5$) near the six-beam orientation, the Burgers vectors of threading edge dislocations (TEDs) were determined using the $\boldsymbol{g}\cdot\boldsymbol{b}$ invisibility criterion. The measured line widths of dislocation images showed excellent quantitative agreement with theoretical predictions based on the extinction distance and $|\boldsymbol{g}\cdot\boldsymbol{b}|$ dependence, confirming that most dislocations observed were $\boldsymbol{a}$-type TEDs with Burgers vectors of $\frac{1}{3}\langle 11\bar{2}0\rangle$, and a small portion with $\frac{2}{3}\langle 11\bar{2}0\rangle$. Since the $\boldsymbol{g}$-vectors $\boldsymbol{g}_1$–$\boldsymbol{g}_5$ used in this work are insensitive to the $\langle 0001\rangle$ component, it is noted that the dislocation analyzed here might possess both $\boldsymbol{c}$ and $\boldsymbol{a}$ components.

These results demonstrate that SR-XRT under multi-beam (n-beam) diffraction provides a powerful and nondestructive technique for dislocation analysis in thick and highly perfect GaN crystals. The method offers a pathway to evaluate dislocation types, densities, and distributions with high spatial resolution and quantitative accuracy, contributing to the development of next-generation GaN substrates and devices. Future work will extend this approach to investigate other defect types and assess strain fields under various multi-beam diffraction geometries.

## Supplementary Material

See Supplementary Figure Fig. S1 for six-beam diffraction XRT images acquired over a wider area and with a fine $\omega$ interval of 0.0004°.

See Supplementary Movie 1 for the Laue diffraction spots observed on the fluorescent screen during the crystal orientation adjustment, continuously achieving the five two-beam diffraction conditions ($\boldsymbol{g}_1$–$\boldsymbol{g}_5$).

See Supplementary Movie 2 for the continuously recorded XRT images during the crystal orientation adjustment shown in Supplementary Movie 1. The Bragg condition for each of the five $\boldsymbol{g}$-vectors ($\boldsymbol{g}_1$–$\boldsymbol{g}_5$) was sequentially satisfied, and five corresponding sets of XRT images were acquired.

## Acknowledgments

This study was financially supported by (1) JSPS KAKENHI Grant No. 20K05355, 23H01872, and 23K17356, Japan, (2) the Iwatani Naoji Foundation, (3) the Iketani Science and Technology Foundation, (4) the Kurata Grants by the Hitachi Global Foundation, (5) the Sumitomo Foundation, and (6) the Murata Science Foundation. The synchrotron XRT observations were performed at SPring-8 with approval from the Japan Radiation Research Institute under proposal Nos. 2022B3055 and 2023A3055. Y.Y.

gratefully acknowledges Professor Dr. J. Matsui for valuable discussions on dynamical XRD theory and the students of Y.T.'s group for assistance with the XRT experiments.

## AUTHOR DECLARATIONS

**Conflict of interest** The authors have no conflicts of interest to disclose.

**Generative AI** Not used in the manuscript preparation process.

### Author contributions

Yongzhao Yao: conceptualization (equal); writing original draft (lead); apparatus and sample preparation (equal); data curation (lead); investigation (lead); writing review and editing (equal); funding acquisition (lead).

Yoshiyuki Tsusaka: conceptualization (equal); apparatus and sample preparation (equal); writing review and editing (equal).

Yukari Ishikawa: writing review and editing (equal).

All authors have approved the manuscript.

**Data Availability** Raw data were generated at the synchrotron facilities SPring-8. The data that support the findings of this study are available within the article and its supplementary material.

**Table I** The values of |$\boldsymbol{g}\cdot\boldsymbol{b}$| for TEDs with a $\boldsymbol{a}$-type $\frac{1}{3}\langle 11\bar{2}0\rangle$ Burgers vectors at $\boldsymbol{g}_1$–$\boldsymbol{g}_5$. The directions of Burgers vectors and $g$-vectors are shown in **Fig. 9**.

| $\lvert\boldsymbol{g}\cdot\boldsymbol{b}\rvert$ | $\boldsymbol{g}_1$ | $\boldsymbol{g}_2$ | $\boldsymbol{g}_3$ | $\boldsymbol{g}_4$ | $\boldsymbol{g}_5$ |
|---|---|---|---|---|---|
| $\boldsymbol{a}_1$ | 1 | 1 | 0 | 1 | 1 |
| $\boldsymbol{a}_2$ | 0 | 1 | 2 | 2 | 1 |
| $\boldsymbol{a}_3$ | 1 | 2 | 2 | 1 | 0 |

## Figure captions

**Fig. 1** (a) Optical system for observing dislocations in thick GaN substrates using synchrotron-radiation X-ray topography based on anomalous transmission near six-beam diffraction conditions; (b) Photograph of the diffraction pattern on the fluorescent screen under six-beam diffraction excitation; (c) Schematic illustration of the X-ray propagation paths and the indexing of the diffraction pattern. The reciprocal lattice vectors $\boldsymbol{g}_1$ to $\boldsymbol{g}_5$ correspond to the wavevectors $\boldsymbol{K}_1$ to $\boldsymbol{K}_5$ shown in panel (a).

**Fig. 2** Integral intensity of the forward-transmitted wave as a function of $\Delta\omega$, the deviation of the incident angle $\omega$ from the six-beam diffraction condition. XRT images obtained with the o-wave using a digital detector are shown for the six-beam diffraction condition and at two half-intensity points.

**Fig. 3** XRT images obtained with the forward-transmitted wave for $\Delta\omega$ ranging from −0.007° to 0.003° in 0.001° intervals. The exact six-beam diffraction condition corresponds to panel (h). Image brightness and contrast were adjusted for clarity.

**Fig. 4** Wide-area XRT image obtained under the six-beam diffraction condition from a 3.5 mm × 3.4 mm region near the substrate center. The image was constructed by tiling 7 × 7 sub-images, each recorded while translating the sample in the X–Y plane parallel to the sample surface.

**Fig. 5** Illustration based on the Ewald sphere showing how five two-beam diffraction conditions, $\boldsymbol{g}_1$–$\boldsymbol{g}_5$, are generated near the six-beam diffraction condition by adjusting the $\psi$ and $\varphi$ angles.

**Fig. 6** Integral intensity of the forward-transmitted wave as a function of $\Delta\omega$, the deviation of the incident angle $\omega$ from the two-beam diffraction condition at $\boldsymbol{g}_3 = 02\bar{2}0$. The two-beam condition at $\boldsymbol{g}_3 = 02\bar{2}0$ was obtained by rotating the $\psi$ angle by 0.01° from the six-beam diffraction condition. XRT images obtained with the o-wave using a digital detector are shown for each of the five two-beam diffraction conditions, $\boldsymbol{g}_1$–$\boldsymbol{g}_5$.

**Fig. 7** XRT images obtained with the forward-transmitted wave under two-beam diffraction conditions: (a) $\boldsymbol{g}_1 = 10\bar{1}0$, (b) $\boldsymbol{g}_2 = 02\bar{2}0$, and (c) $\boldsymbol{g}_3 = \bar{1}100$. For each condition, a sequence of images was recorded while rocking the $\omega$ angle from −0.004° to 0.004°.

**Fig. 8** Line width of dislocations observed under kinematical contrast. Pixel brightness profiles and their Gaussian fittings around (a) dislocation 1 and (b) dislocation 2. (c) XRT image showing the two dislocations and the path along which the line profile was measured.

**Fig. 9** Five $\boldsymbol{g}$-vectors used for two-beam diffraction, shown in reciprocal space. (b) Crystallographic orientation of hexagonal GaN as defined in this work. Note that Fig. 9 is viewed from the Ga-face, whereas Figs. 1(b) and 1(c) are viewed from the N-face.

**Fig. 1 (rotated by 90°)**

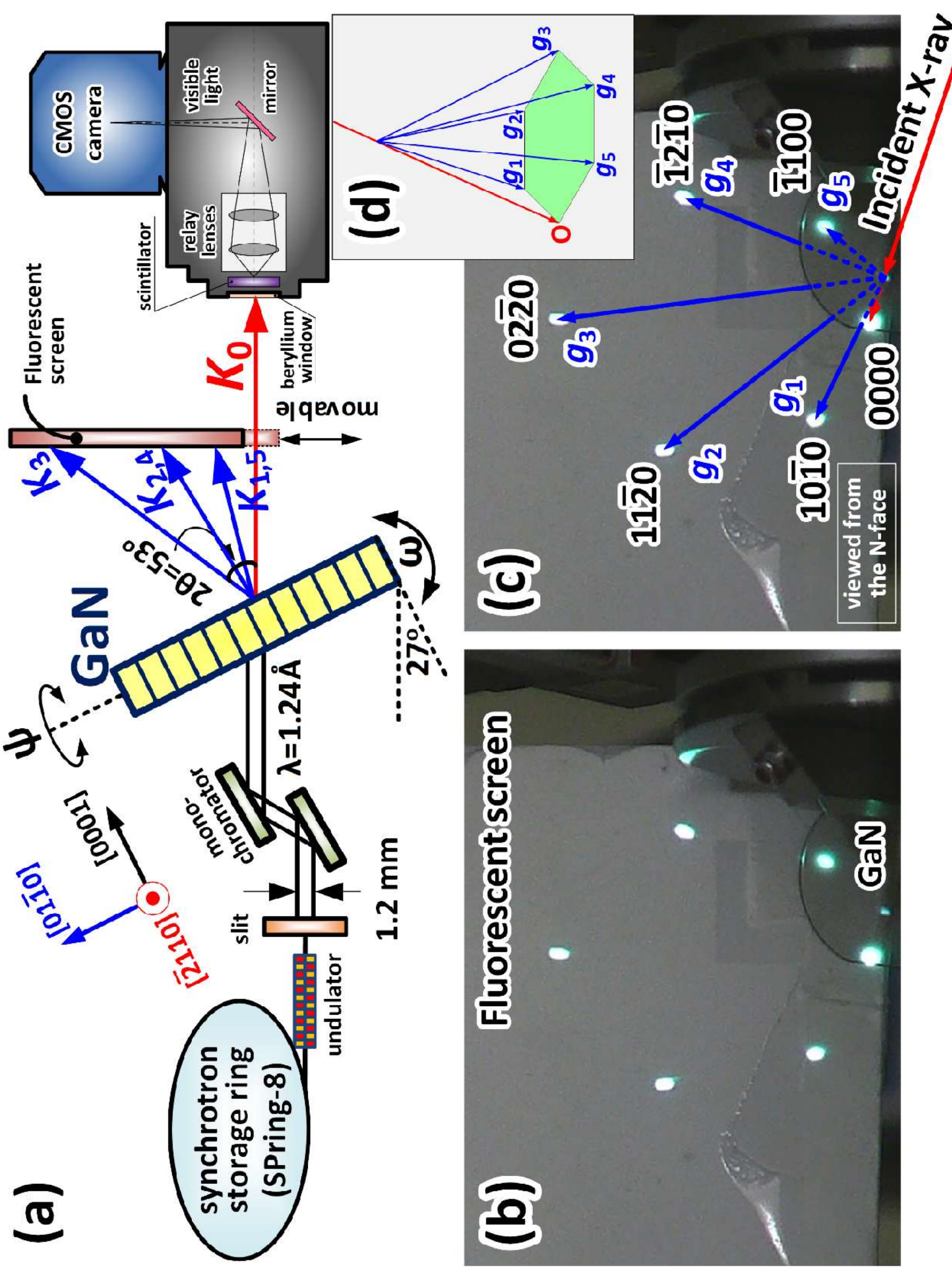

**Fig. 2 (rotated by 90°)**

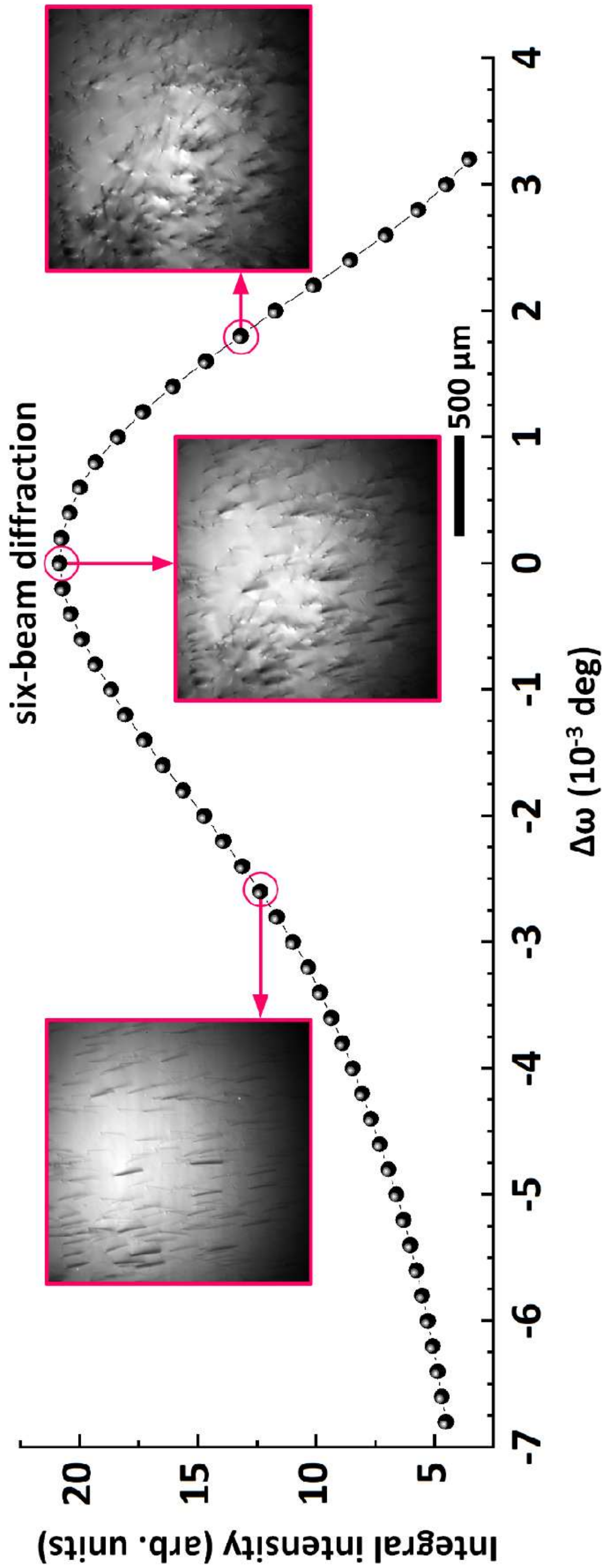

**Fig. 3 (rotated by 90°)**

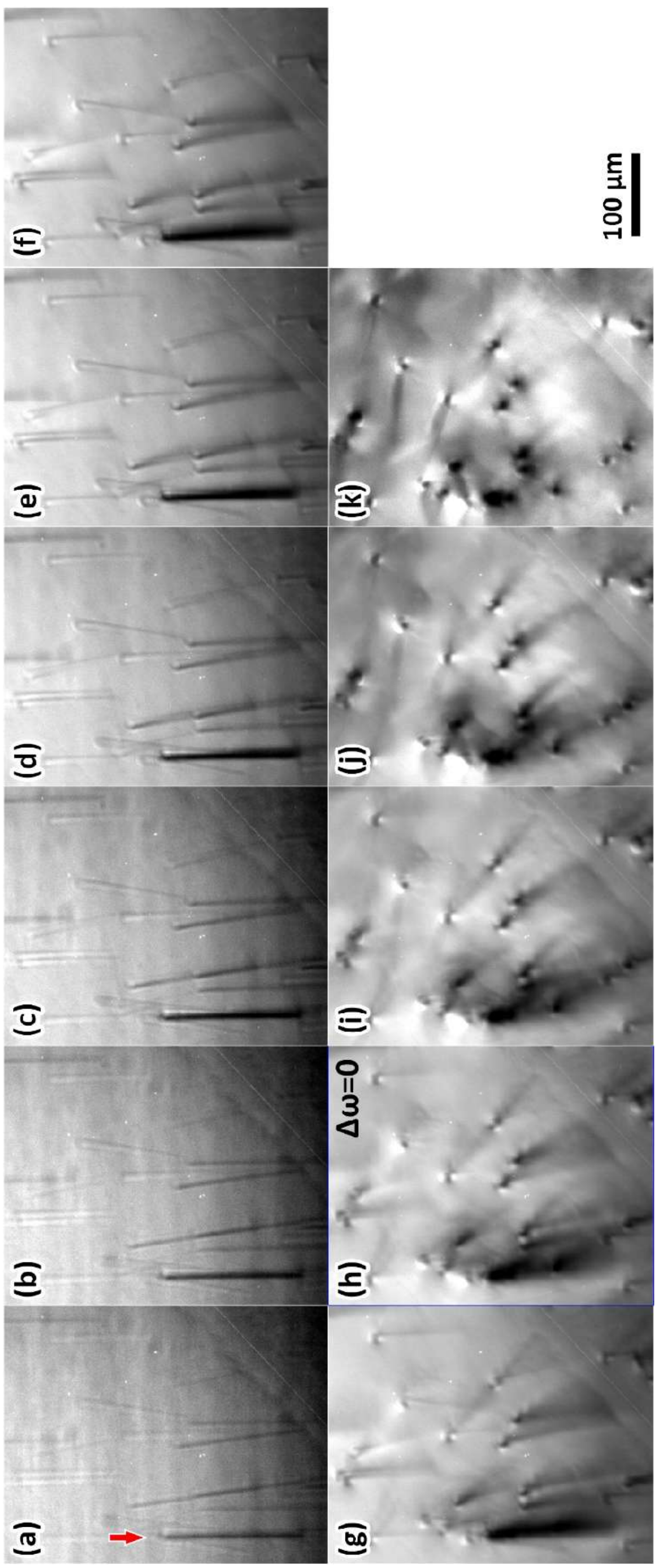

**Fig. 4**

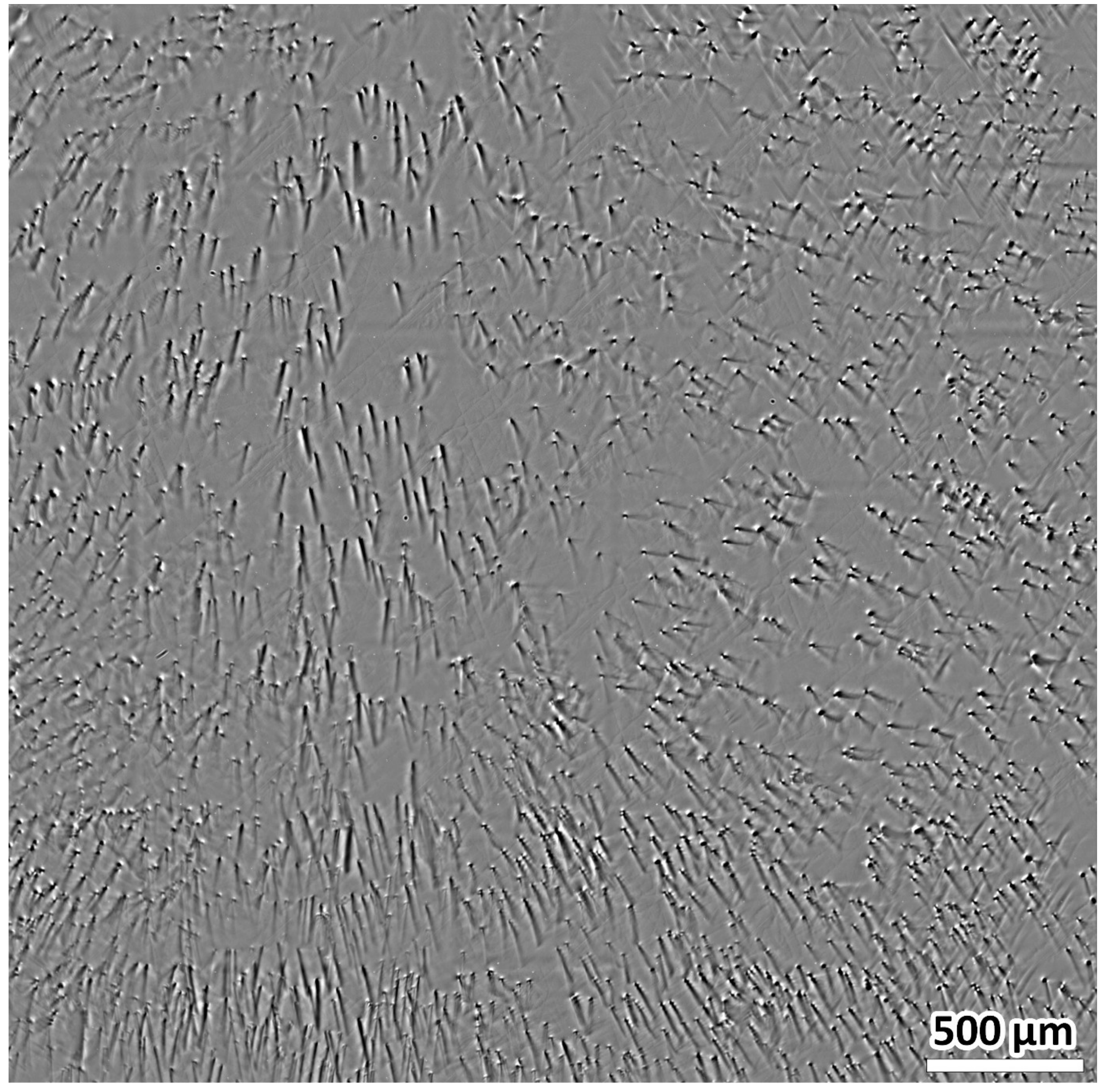

**Fig. 5 (rotated by 90°)**

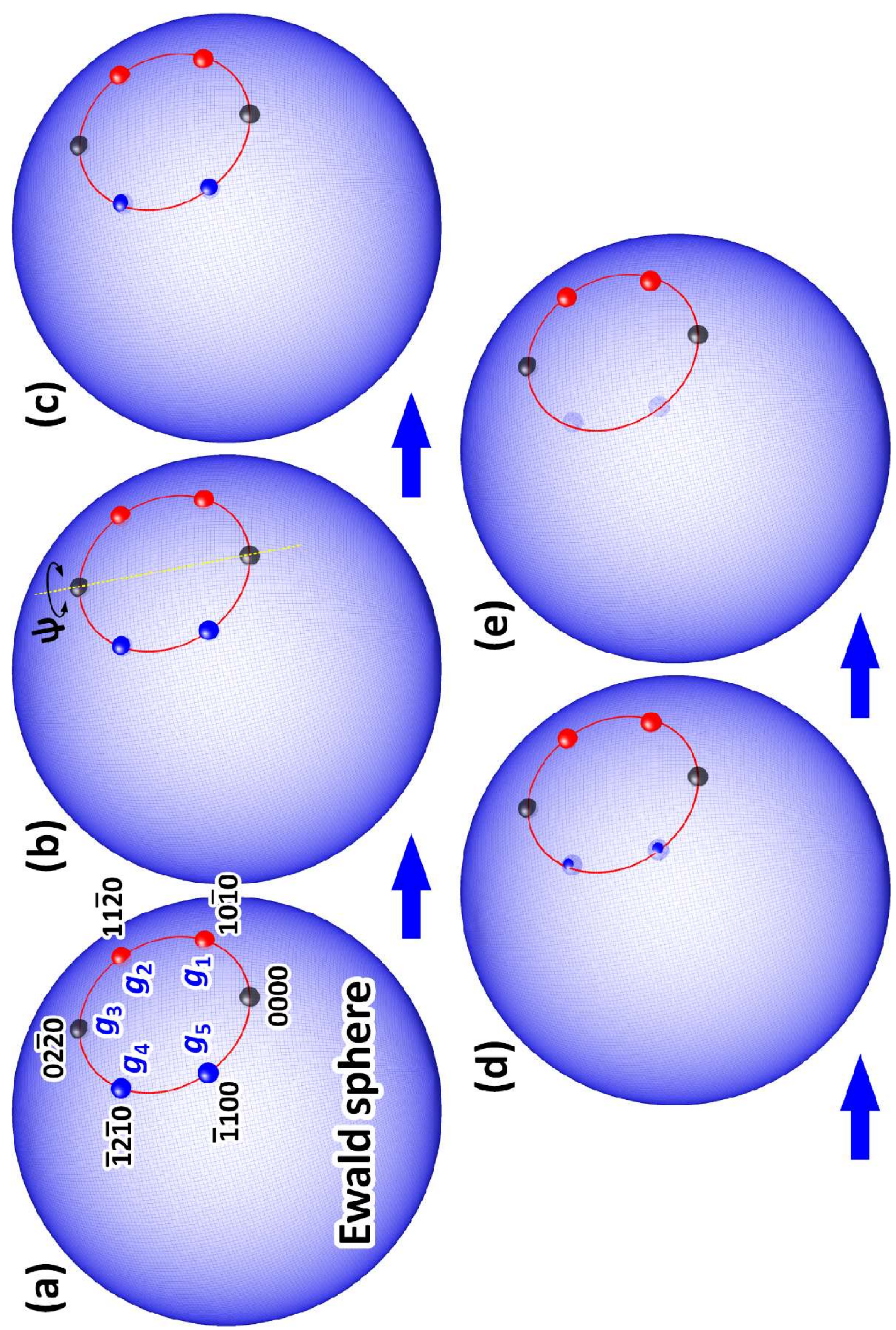

**Fig. 6 (rotated by 90°)**

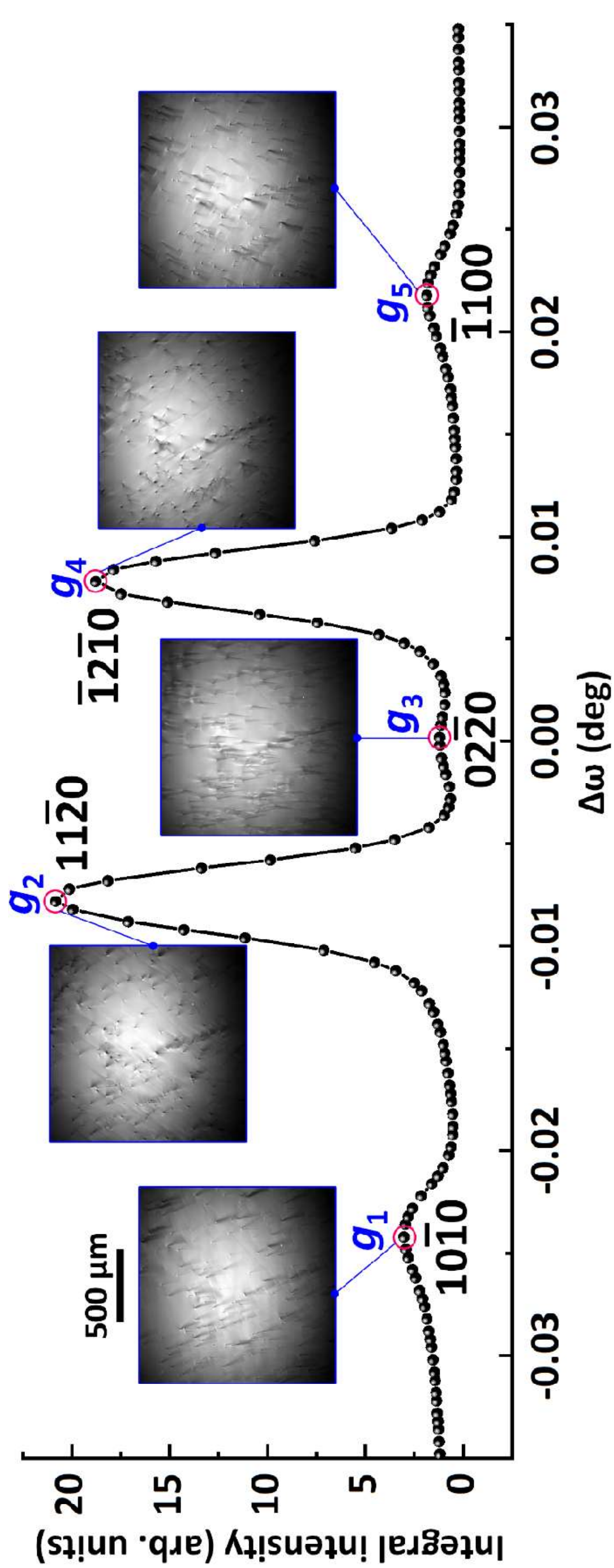

## Fig. 7 (rotated by 90°)

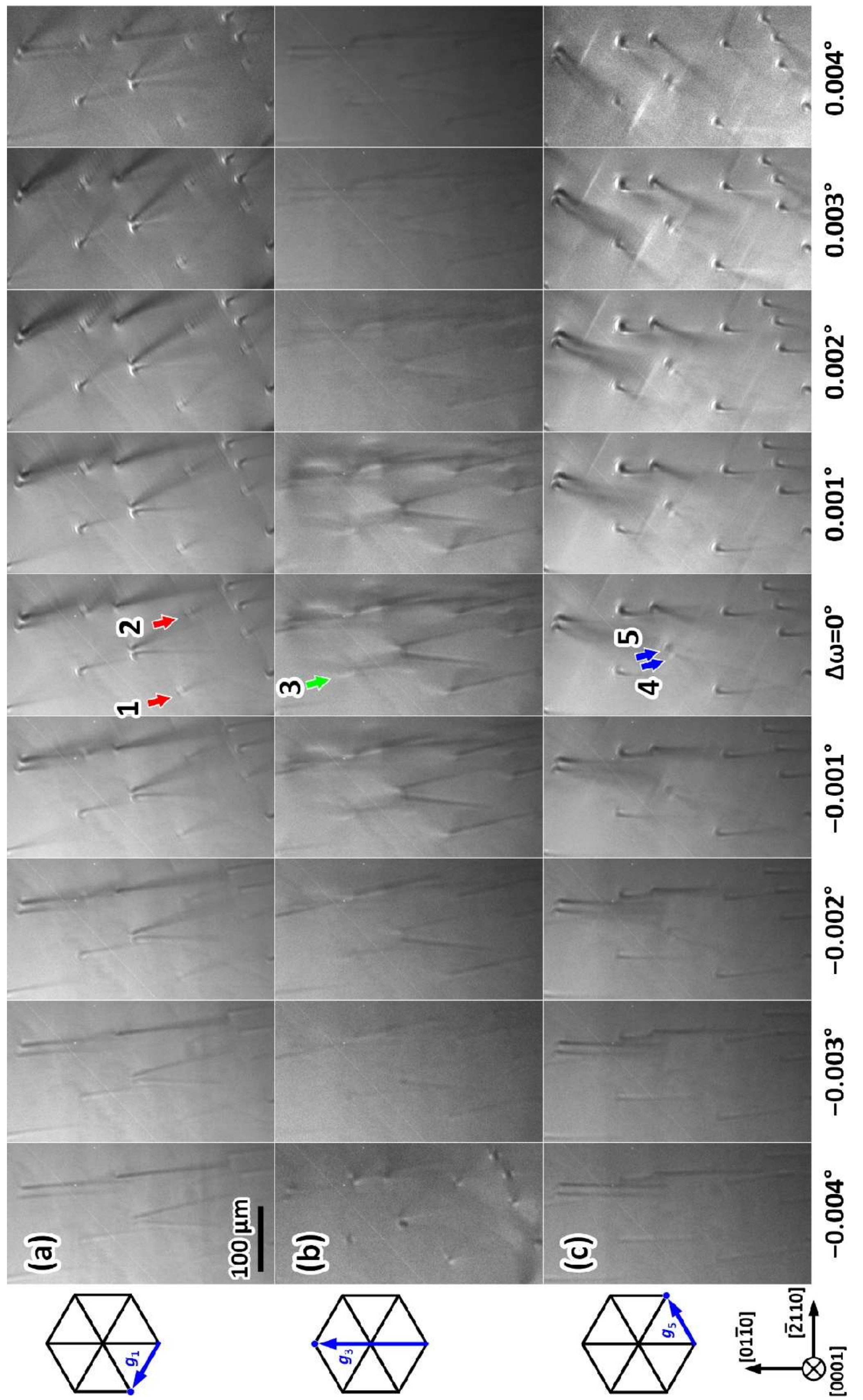

**Fig. 8 (rotated by 90°)**

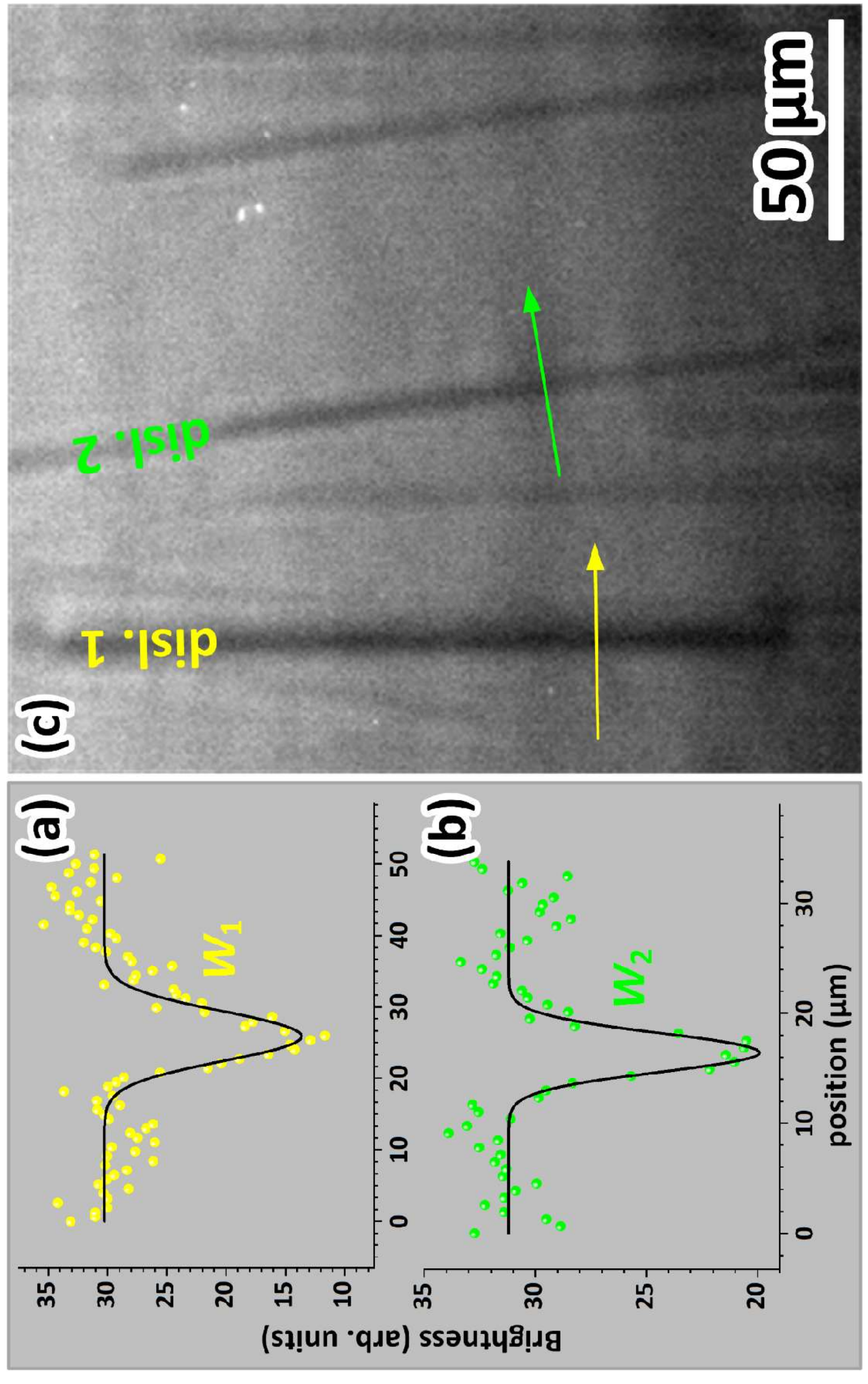

**Fig. 9 (rotated by 90°)**

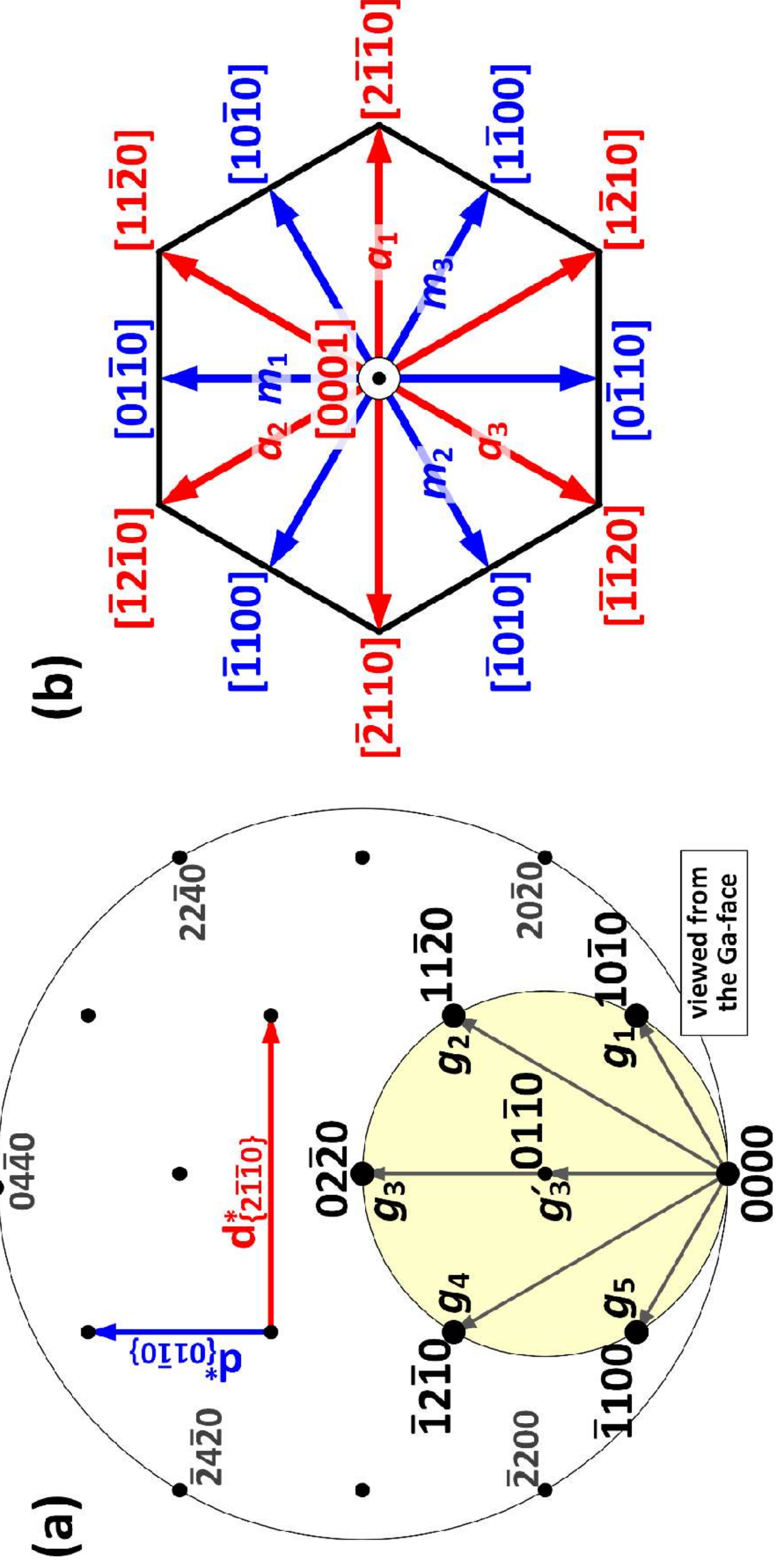

**Fig. S1**

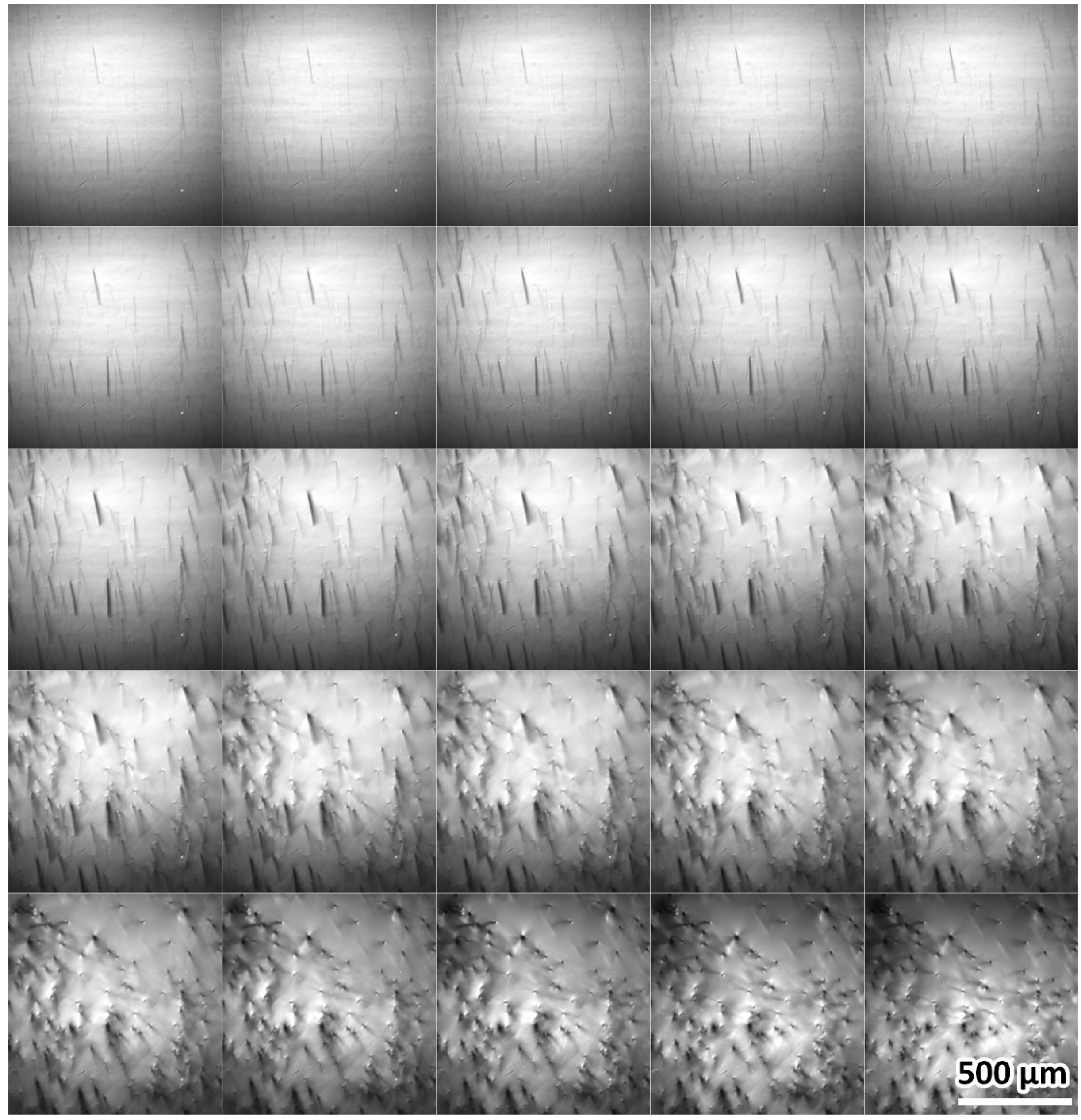